\documentstyle[12pt,twoside,fleqn,espcrc1,epsfig]{article}

\title{Energy Loss of Hard Partons in Nuclear Matter}

\author{Urs Achim Wiedemann\address{
Theory Division, CERN, CH-1211 Geneva 23}}

\begin{document}
\maketitle

\begin{abstract}
We report on recent calculations of the medium-induced gluon radiation 
off hard partons. The employed path-integral formalism reduces in
limiting cases to the main ``jet quenching'' results of the existing 
literature. Moreover, it describes destructive interference effects between
medium-independent and medium-induced radiation amplitudes. These
affect the angular distribution and $L$-dependence of the 
medium-induced gluon bremsstrahlung spectrum significantly.
\end{abstract}

\section{Introduction}

At this conference, the PHENIX and STAR Collaborations have 
presented hadronic transverse momentum spectra in central
Au+Au collisions at $\sqrt{s_{\rm NN}}=130$ GeV\cite{Z01,H01}. 
In comparison to rescaled $p+\bar{p}$ UA1 reference data,
these spectra show a depletion at high transverse momentum 
($2$ GeV $< p_\perp < 6$ GeV). This may be the first experimental 
indication for jet quenching~\cite{WLP01},
predicted by Gyulassy and Wang~\cite{WG91} a decade ago.

In general, hadronic transverse momentum spectra contain both a
soft and a hard physics component. The hard component, which
dominates at sufficiently high $p_\perp$, can be calculated by 
convoluting the incoming parton distributions with partonic cross 
sections. Parton fragmentation functions parametrize how the  
partons produced in hard processes translate into final state hadrons. 
Thus, the origin of any medium-dependence of high-$p_\perp$ transverse 
momentum spectra has to be traced back to either i) nuclear modifications
of the initial state (e.g. nuclear shadowing) or ii) medium-induced 
changes of the final state parton fragmentation. While 
information about nuclear shadowing can be obtained from other
observables (e.g. DIS in e-A), the study of the
medium-dependence of parton fragmentation functions $D_q^h(z)$ is 
still in its infancy. Here, we review recent calculations
~\cite{Z96,BDMPS97,BDMS-Zak,WG99,BDMS99,W00qcd,GLV99} of the
radiative energy loss off hard partons which aim at describing
the medium-induced shift 
in the energy of the fragmenting parton. For a complementary approach,
see Ref.~\cite{GW01}.
\vspace{-1.0cm}
\begin{center}
\begin{figure}[h]
\centerline{\epsfig{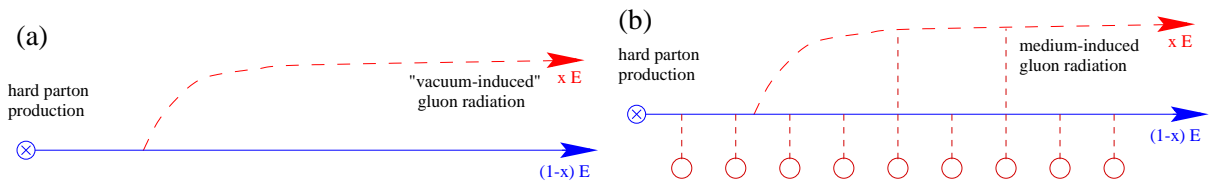}}
\vspace{-1.0cm}
\caption{The fragmentation of a hard quark via gluon radiation
(a) outside and (b) inside a nuclear environment.
}\label{fig1}
\end{figure}
\end{center}
%
\section{Medium-induced Gluon radiation}
Fig.~\ref{fig1}(a) shows a typical building block of the
DGLAP evolution equation. This gluon radiation shifts the
energy of the quark by a factor $1-x$. To
describe medium-induced changes of such partonic fragmentation
patterns, we consider multiple scattering contributions, 
see Fig.~\ref{fig1}(b).
For the Gyulassy-Wang model which mimicks the nuclear medium by a 
set of static coloured scattering potentials, these scattering
contributions were calculated recently by several 
groups~\cite{Z96,BDMPS97,BDMS-Zak,WG99,BDMS99,W00qcd,GLV99}. All 
published results are limiting cases of the following expression 
for the medium-induced gluon radiation cross section off a hard 
quark~\cite{W00qcd}:
\begin{eqnarray}
  &&{d^3\sigma\over d(\ln x)\, d{\bf k}_\perp}
  = {\alpha_s\over (2\pi)^2}\, {1\over \omega^2}\,
    N_C\, C_F\, 
    2{\rm Re} \int_0^{\infty} dy_l  
  \int_{y_l}^{\infty} d\bar{y}_l\, 
  e^{-\epsilon |y_l|\, -\epsilon |\bar{y}_l|}
  \nonumber \\
  && \qquad \times 
  \int d{\bf u}\,   e^{-i{\bf k}_\perp\cdot{\bf u}}   \, 
  e^{ -\frac{1}{2} \int_{\bar{y}_l}^{\infty} d\xi\, n(\xi)\, 
    \sigma({\bf u}) }\,
  {\partial \over \partial {\bf y}}\cdot 
  {\partial \over \partial {\bf u}}\, 
  {\cal K}({\bf y}=0,y_l; {\bf u},\bar{y}_l|\omega) \, .
    \label{2.1}
\end{eqnarray}
Here, $x$ denotes the energy fraction of the hard quark,
carried away by the gluon, and ${\bf k}_\perp$ its transverse
momentum. Information about the nuclear medium enters via the
combination $n(\xi)\, \sigma({\bf r})$, where $n(\xi)$ determines
the density of scattering centers in the medium, and the dipole
cross section $\sigma({\bf r}) \propto \int d{\bf q}_\perp\, 
|a_0({\bf q}_\perp)|^2 \left( 1 - e^{-i{\bf q}_\perp\cdot {\bf r}}\right)$
contains configuration space information on the 
strength of a single elastic differential scattering cross 
section $|a_0({\bf q}_\perp)|^2$. 
${\cal K}$ denotes the two-dimensional path-integral~\cite{Z96}
\begin{equation}
 {\cal K}({\bf r}(y_l),y_l;{\bf r}(\bar{y}_l),\bar{y}_l|\omega)
 = \int {\cal D}{\bf r}
   \exp\left[ \int_{y_l}^{\bar{y}_l} d\xi
        \left(i\frac{\omega}{2} \dot{\bf r}^2
          - \frac{1}{2}  n(\xi) \sigma\left({\bf r}\right) \right)
                      \right]\, ,
  \label{2.2}
\end{equation}
%

\section{Opacity Expansion}
In relativistic heavy ion collisions, hard partons typically
propagate over distances $L = 1-10 \times \lambda_{\rm mfp}$ 
inside the excited nuclear medium. To access this region, we
expand the radiation spectrum (\ref{2.1}) in powers of
opacity $\left(\alpha_s \int_0^L d\xi\, n(\xi)\right)^N$ which
is related to an expansion in the number of scattering
centers. For arbitrary $N$, it can be shown that the radiation 
spectrum (\ref{2.1}) interpolates between the classically expected 
$N$-fold scattering results in both the coherent ($L\to 0$)
and incoherent ($L\to \infty$) limiting cases~\cite{W00qcd}.
For example, up to first order~\cite{W00qcd} 
\begin{eqnarray}
  &&{d^3\sigma(N=0,1)\over d(\ln x)\, d{\bf k}_\perp}
  = {\alpha_s\over \pi^2}\, N_C\, C_F\,
    \left( \frac{1}{{\bf k}^2_\perp} +
    \frac{1}{2\,\omega^2}\, 
    \int_{\Sigma_1}
    \frac{-{\bf k}_\perp\cdot {\bf q}_\perp}{Q\, Q_1}\,
    n_0\, \frac{LQ_1 - \sin(LQ_1)}{Q_1} \right)
    \nonumber\\ 
  && \Longrightarrow_{\lim_{L\to\infty}} {\alpha_s\over \pi^2}\, N_C\, C_F
    \left[ 
    (1-w_1)\, H({\bf k}_\perp) + n_0\, L\, \int_{{\bf q}_1} 
    \left[ H({\bf k}_\perp + {\bf q}_{\perp}) +
     R({\bf k}_\perp,{\bf q}_{\perp}) \right]
                  \right]\, .
  \label{2b.1}
\end{eqnarray}
Here, $Q = \frac{{\bf k}_\perp^2}{2\omega}$, 
$Q_1 = \frac{({\bf k}_\perp + {\bf q}_\perp)^2}{2\omega}$ are
transverse energies, the ``hard'' term ${\alpha_s\over \pi^2}\, N_C\, C_F\, 
H({\bf k}_\perp)$, $H({\bf k}_\perp) = \frac{1}{{\bf k}^2_\perp}$,
is the medium-independent contribution of Fig.~\ref{fig1}(a) and
$\int_{{\bf q}_1}$ is a shorthand for integrating over the elastic
scattering cross section. Eq. (\ref{2b.1})
is the incoherent ``classical'' parton cascade limit of (\ref{2.1}):
the medium-independent hard term $H({\bf k}_\perp)$ is reduced by
the probability $w_1$ of an additional scattering. With
the probability for this scattering, the hard term is shifted in
transverse momentum to $H({\bf k}_\perp + {\bf q}_{\perp})$, and
the medium-induced gluon radiation off the hard quark leads to 
a Gunion-Bertsch radiation term $R({\bf k}_\perp,{\bf q}_{\perp}) =
\frac{{\bf q}_{\perp}^2}{{\bf k}_\perp^2\, 
({\bf k}_\perp+{\bf q}_{\perp})^2}$.
The ${\bf k}_\perp$-integral of 
the medium-dependent part of (\ref{2b.1}) is the GLV-radiation
cross section~\cite{GLV99}. This lowest order opacity approximation
was compared to RHIC data at this conference~\cite{WLP01}.
%
\begin{center}
\begin{figure}[t]
\centerline{\epsfig{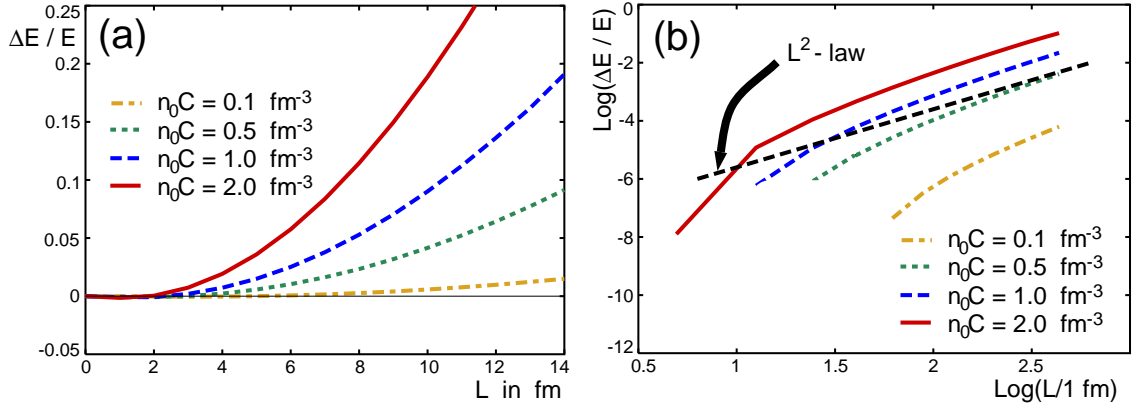}}
\vspace{-1.0cm}
\caption{(a) Medium-induced energy loss for a quark of $E = 100$ GeV 
as a function of the
in-medium path length $L$. (b) The double logarithmic presentation
of (a) indicates logarithmic deviations from the BDMPS-$L^2$-law.
}\label{fig2}
\end{figure}
\end{center}
%
\vspace{-1cm}
\section{Dipole Approximation}
For large in-medium path length $L \gg \lambda_{\rm mfp}$, the
integrand of the radiation cross section (\ref{2.1}) is dominated 
by small transverse distances ${\bf u}$ for which 
$\sigma({\bf u}) \approx C\, {\bf u}^2$ up to logarithmic
accuracy. In this quadratic approximation, the path-integral
${\cal K}$ is that of a harmonic oscillator, and (\ref{2.1})
can be reduced to a two-dimensional integral which has to
be evaluated numerically~\cite{Z96,W00qcd}. The medium dependence enters then
via the rescattering parameter $n_0\, C$ which parametrizes the
average squared transverse momentum picked up by a hard parton
per unit pathlength, $n_0\, C = \frac{\langle {\bf q}_\perp^2\rangle}{L}$.
This is a measure of the transverse colour field strength seen by
a hard parton~\cite{W00qcd}. In cold nuclear matter, 0.1 fm$^{-3}$ 
$< n_0\, C <$ 0.5 fm$^{-3}$ while in heavy ion collisions, one 
expects $n_0\, C > $ 1.0 fm$^{-3}$~\cite{BDMPS97,W00qcd}.
In what follows, we consider the 
medium-induced deviation of the radiation spectrum, 
$\sigma_{\rm med}(n_0\, C) = \sigma(n_0\, C) - \sigma(n_0\, C = 0)$. The
radiative energy loss $\Delta E$ is
\begin{equation}
  \frac{\Delta E}{E}(n_0\, C,L,E,\chi) = \int_0^1 dx\, x\,
   \int_{|{\bf k}_\perp|\leq \chi\omega}  d{\bf k}_\perp
        \frac{d^3\sigma_{\rm med}}{d({\rm ln}x)\,d{\bf
        k}_\perp}\, . 
 \label{3.1}
\end{equation}
For $\chi = 1$, the ${\bf k}_\perp$-integral goes up to the
kinematical boundary ${\bf k}_\perp = \omega$ and (\ref{3.1})
gives the total radiative energy loss. For typical spatial extensions
in heavy ion collisions, $L < 10$ fm, we find that the radiative 
energy loss is very sensitive to the rescattering properties
$n_0\, C$ of the medium, see  Fig.~\ref{fig2}. Its
$L$-dependence satisfies approximately the $L^2$-law 
discovered by BDMPS. However, the destructive interference between hard
and medium-induced radiation, leads to a delayed onset
of any radiative energy loss: $\Delta E$ in Fig.~\ref{fig2}
turns negative for $L < 2$ fm. This increases the fraction of
hard partons whose in-medium path length is too small to result
in a significant medium-dependence (``corona effect'').  
%
\begin{center}
\begin{figure}[t]
\centerline{\epsfig{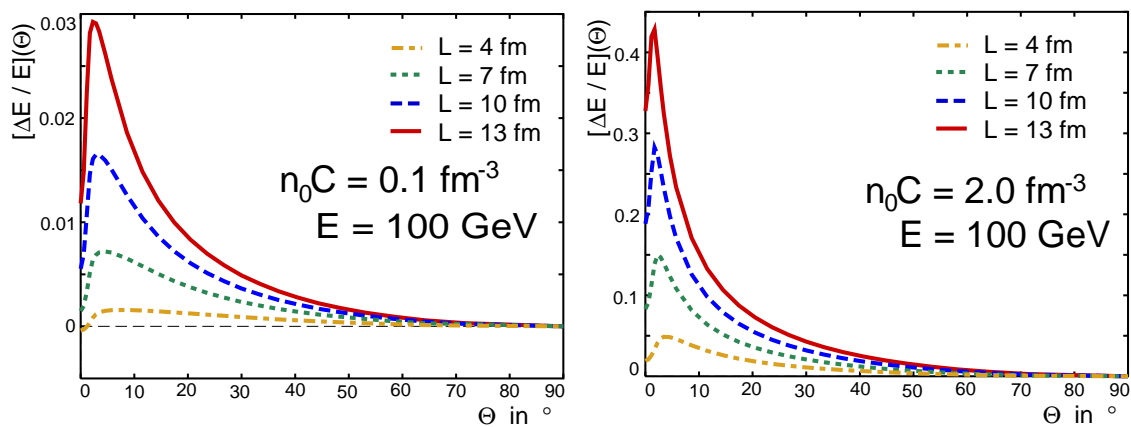}}
\vspace{-1.0cm}
\caption{The medium-induced deviation of the energy radiated 
outside a finite jet opening angle $\Theta$ has a maximum at
finite angle. 
}\label{fig3}
\end{figure}
\end{center}
%
\vspace{-1.0cm}
The $\chi$-dependence of (\ref{3.1}) translates into an opening angle 
and allows to calculate radiative energy loss inside a finite jet
cone. While the 
total energy radiated outside a finite angle $\Theta$ decreases
monotonously with increasing opening angle $\Theta$, this is not
the case for the medium-induced deviation obtained from (\ref{3.1})
and $\sigma_{\rm med}(n_0\, C) = \sigma(n_0\, C) - \sigma(n_0\, C = 0)$.
Medium-induced multiple scattering broadens the medium-independent
term $\sigma(n_0\, C = 0) \propto \frac{1}{{\bf k}_\perp^2}
\to \frac{1}{({\bf k}_\perp + {\bf q}_\perp)^2}$. This leads to a 
maximum of $\Delta E (\Theta)$ for finite opening angle. As a
consequence, the ${\bf k}_\perp$-integrated energy loss 
expressions $\Delta E (\Theta=0)$ presented in previous works 
are not necessarily upper bounds for the energy lost outside a 
realistic jet cone.



\begin{thebibliography}{9}
\bibitem{Z01}
 W.A. Zajc and A. Milov [PHENIX Collaboration], Talks at Quark Matter 2001.
\bibitem{H01}
 J. Harris and J.C. Dunlop [STAR Collaboration], Talks at Quark Matter 2001.
\bibitem{WLP01}
 X.N. Wang and P. Levai, Talks at Quark Matter 2001.
\bibitem{WG91}
 X.-N. Wang and M. Gyulassy,  Phys. Rev. {\bf D44} (1991) 3501; 
 M. Gyulassy and X.-N. Wang, Nucl. Phys. {\bf B420}
 (1994) 583.
\bibitem{Z96} 
 B.G. Zakharov, JETP Letters {\bf 63} (1996) 952; ibidem
 {\bf 65} (1997) 615;
 {\bf 70} (1999) 176. 
\bibitem{BDMPS97}
 R. Baier, Y.L. Dokshitzer, A.H. Mueller, S. Peign\'e and D. Schiff,
 Nucl. Phys. {\bf B483} (1997) 291;
 ibidem {\bf B484} (1997) 265.
\bibitem{BDMS-Zak}
 R. Baier, Y.L. Dokshitzer, A.H. Mueller and D. Schiff,
 Nucl. Phys. {\bf B531} (1998) 403.
\bibitem{WG99} U.A. Wiedemann and M. Gyulassy, Nucl. Phys.
 {\bf B560} (1999) 345.
\bibitem{BDMS99}
 R. Baier, Y.L. Dokshitzer, A.H. Mueller and D. Schiff,
 Phys. Rev. {\bf  C60} (1999) 064902.
\bibitem{W00qcd}
 U.A. Wiedemann, Nucl. Phys. {\bf B582} (2000) 409; 
Nucl. Phys. {\bf B588} (2000) 303; hep-ph/0008241.
\bibitem{GLV99} M. Gyulassy, P. Levai and I. Vitev, 
 Nucl. Phys. {\bf B571} (2000) 197, nucl-th/0005032 and nucl-th/0006010.
\bibitem{GW01}
 X.F. Guo and X.N. Wang, Phys. Rev. Lett. {\bf 85} (2000) 3591 and
 hep-ph/0102230.  
\end{thebibliography}
\end{document}